\documentclass{aastex62}

\graphicspath{{figures/}}

\shorttitle{Kinematics of NGC~1261}
\shortauthors{Raso et al.}

\begin{document}

\title{A kinematic view of NGC 1261: structural parameters, internal dispersion, absolute proper motion and Blue Straggler Stars}

\correspondingauthor{Silvia Raso}
\email{silvia.raso@inaf.it} \email{silvia.raso2@unibo.it}

\author[0000-0003-4592-1236]{Silvia Raso} \affil{Istituto Nazionale di Astrofisica (INAF), Osservatorio di Astrofisica e Scienza dello Spazio di Bologna, Via Gobetti 93/3, Bologna I-40129, Italy} \affil{Dipartimento di Fisica e Astronomia, Universit\`a di Bologna, Via Gobetti 93/2, Bologna I-40129, Italy} 

\author[0000-0001-9673-7397]{Mattia Libralato} \affil{Space Telescope Science Institute, 3700 San Martin Drive, Baltimore, MD 21218, USA}

\author[0000-0003-3858-637X]{Andrea Bellini} \affil{Space Telescope Science Institute, 3700 San Martin Drive, Baltimore, MD 21218, USA}

\author[0000-0002-2165-8528]{Francesco R. Ferraro} \affil{Dipartimento di Fisica e Astronomia, Universit\`a di Bologna, Via Gobetti 93/2, Bologna I-40129, Italy}  \affil{Istituto Nazionale di Astrofisica (INAF), Osservatorio di Astrofisica e Scienza dello Spazio di Bologna, Via Gobetti 93/3, Bologna I-40129, Italy}

\author[0000-0001-5613-4938]{Barbara Lanzoni} \affil{Dipartimento di Fisica e Astronomia, Universit\`a di Bologna, Via Gobetti 93/2, Bologna I-40129, Italy}  \affil{Istituto Nazionale di Astrofisica (INAF), Osservatorio di Astrofisica e Scienza dello Spazio di Bologna, Via Gobetti 93/3, Bologna I-40129, Italy}

\author[0000-0002-5038-3914]{Mario Cadelano} \affil{Dipartimento di Fisica e Astronomia, Universit\`a di Bologna, Via Gobetti 93/2, Bologna I-40129, Italy}  \affil{Istituto Nazionale di Astrofisica (INAF), Osservatorio di Astrofisica e Scienza dello Spazio di Bologna, Via Gobetti 93/3, Bologna I-40129, Italy}

\author[0000-0002-7104-2107]{Cristina Pallanca} \affil{Dipartimento di Fisica e Astronomia, Universit\`a di Bologna, Via Gobetti 93/2, Bologna I-40129, Italy}  \affil{Istituto Nazionale di Astrofisica (INAF), Osservatorio di Astrofisica e Scienza dello Spazio di Bologna, Via Gobetti 93/3, Bologna I-40129, Italy}

\author[0000-0003-4237-4601]{Emanuele Dalessandro} \affil{Istituto Nazionale di Astrofisica (INAF), Osservatorio di Astrofisica e Scienza dello Spazio di Bologna, Via Gobetti 93/3, Bologna I-40129, Italy}

\author[0000-0002-9937-6387]{Giampaolo Piotto}
\affil{Dipartimento di Fisica e Astronomia ``Galileo Galilei'', Universit\`a degli Studi di Padova, Vicolo dell’Osservatorio 3, I-35122 Padova, Italy} \affil{Istituto Nazionale di Astrofisica (INAF), Osservatorio Astronomico di Padova, Vicolo dell’Osservatorio 5, I-35122 Padova, Italy}

\author[0000-0003-2861-3995]{Jay Anderson} \affil{Space Telescope Science Institute, 3700 San Martin Drive, Baltimore, MD 21218, USA}

\author[0000-0001-8368-0221]{Sangmo Tony Sohn}
\affil{Space Telescope Science Institute, 3700 San Martin Drive, Baltimore, MD 21218, USA}

\begin{abstract}

We constructed a \textit{Hubble Space Telescope} (\textit{HST}) astro-photometric catalog of the central region of the Galactic globular cluster NGC~1261. This catalog, complemented with \textit{Gaia} DR2 data sampling the external regions, has been used to estimate the structural parameters of the system (i.e., core, half-mass, tidal radii and concentration) from its resolved star density profile. We computed high-precision proper motions thanks to multi-epoch \textit{HST} data and derived the cluster velocity dispersion profile in the plane of the sky for the innermost region, finding that the system is isotropic. The combination with line-of-sight information collected from spectroscopy in the external regions provided us with the cluster velocity dispersion profile along the entire radial extension. We also measured the absolute proper motion of NGC~1261 using a few background galaxies as a reference. The radial distribution of the Blue Straggler Star population shows that the cluster is in a low/intermediate phase of dynamical evolution.

\end{abstract}

\keywords{Globular Clusters: individual: NGC~1261; Proper motions; Stars: kinematics and dynamics; Techniques: photometric}

\section{Introduction} \label{sec:intro}

Globular clusters (GCs) are fundamental laboratories to test stellar dynamics. Their internal dynamical evolution is mainly led by two body encounters, during which stars exchange energy and progressively lose memory of their initial conditions (e.g., \citealt{binneytremaine87, meylanheggie97}). In general, in GCs the relaxation timescale is shorter than the age of the system, and therefore the effects of dynamics can be directly observed.

In this respect, we are carrying on a long-term project aimed at the accurate characterization of the internal structure, kinematics and stage of dynamical evolution of GCs, by using (1) star density profiles (\citealt{lanzoni07a,lanzoni07b,lanzoni07c, lanzoni10, lanzoni19, miocchi13}) to derive the cluster structural parameters, (2) velocity dispersion profiles from line-of-sight (LOS) velocities and proper motions (PMs) of individual stars (see \citealt{lanzoni13, lanzoni18a, lanzoni18b, ferraro18mikis} and \citealt{bellini15, libralato18}, respectively for LOS velocities and PMs), and (3) the properties of Blue Stragglers Stars (BSSs), a particular class of exotic objects that have been found  to be powerful tracers of the internal dynamical evolution of stellar systems (see \citealt{ferraro09, ferraro12, ferraro18treasury, ferraro19, lanzoni16, dalessandro13, beccari19}).

In particular: (1) we are using number counts, instead of surface brightness distributions, to determine the cluster gravitational centers and the structural and kinematical parameters. This approach allows us to avoid the so-called ``shot-noise bias'' that is known to affect the surface brightness distribution in case of resolved stellar populations. In fact the presence of a few luminous stars can significantly alter the correct localization of the center of gravity and the shape of the star density profile (e.g., \citealt{calzetti93, lugger95, montegriffo95}).

(2) We are using a combination of LOS velocities and PMs of individual stars to derive the velocity dispersion profile of star clusters (\citealt{lanzoni13, lanzoni18a, lanzoni18b, ferraro18mikis, bellini15, bellini18, libralato18}). This approach has been found to bring much more solid pieces of information in the case of resolved stellar populations than present-day integrated-light spectroscopy, (i.e., measuring the line broadening and Doppler shift from integrated light spectra), which is again prone to a severe ``shot noise bias'', since the spectra can be dominated by the light of just a few bright stars (e.g., \citealt{dubath97}).  

(3) In order to complete the physical characterization of high density stellar systems, we are using the observational properties of BSSs, which are powerful tracers of GC internal dynamics (\citealt{ferraro12, ferraro18treasury}). BSSs are a peculiar stellar population, placed in the color-magnitude diagram (CMD) along an extension of the main sequence (MS), at brighter and bluer magnitudes with respect to the turn-off (TO; see \citealt{ferraro92, ferraro93, ferraro97, ferraro04, ferraro18treasury, piotto04, lanzoni07a, lanzoni07b, dalessandro08, moretti08, beccari11, beccari12, leigh11, simunovicpuzia16}). Their position in the CMD and direct measurements suggest that they are more massive than the TO mass (\citealt{shara97, gilliland98, demarco05, ferraro06, fiorentino14, baldwin16, libralato18, libralato19, raso19}), and for this reason they are subject to dynamical friction, which progressively makes them sink towards the cluster center (e.g., \citealt{mapelli04, mapelli06}). Therefore, the BSS radial distribution is expected to be modified as a function of the cluster dynamical age, and, based on this, \citet{ferraro12} first defined the concept of the so-called ``dynamical clock''. More recently, \citet{alessandrini16, lanzoni16, ferraro18treasury, ferraro19} refined this tool and applied it to rank stellar systems in the Galaxy and in the Large Magellanic Cloud as a function of their dynamical age (see also \citealt{raso17, ferraro19, li19, singhyadav19, sollimaferraro19}).

The present paper is devoted to compute the structural, kinematical and dynamical properties of the Galactic GC NGC~1261. This is an intermediate-metallicity GC, with very low extinction ([Fe/H]=$-$1.27 and E($B-V$)=0.01, \citealt{h96}, 2010 edition). It is a halo GC, with a distance modulus $(m-M)_0 = 15.98$ (\citealt{ferraro99a}), corresponding to a distance from the Sun of 15.7 kpc. It is located farther from the Sun with respect to GCs studied in most of the previous works dealing with internal kinematics from \textit{HST} PMs, which all have distances in the range 5$-$10~kpc (\citealt{b14, bellini15, bellini18, libralato18, libralato19}). However, its extremely low extinction and its expected low contamination from field stars (due to its low Galactic latitude, B=$-52.12^{\circ}$, and its relatively large distance from the Galactic center, D=18.1~kpc, \citealt{h96}), make it a suitable candidate for PM studies and to test the validity of PM measurements in a less favourable (more distant) case. 
\citet{simunovic14} proposed the presence of an interesting BSS population, with two separate sequences in the CMD. 
In addition, NGC~1261 is suspected to have an extra-tidal structure, in the form of an outer envelope (\citealt{kuzma18, shipp18}), and to be associated to the \textit{Gaia}-Enceladus structure (\citealt{massari19}).

The paper is organized as follows: in Section~\ref{sec:datared} we describe the dataset and the data reduction procedure; in Section~\ref{sec:struct} we derive the structural parameters of the cluster from resolved stellar counts; in Section~\ref{sec:kin} we construct and study the velocity dispersion profiles from PMs; in Section~\ref{sec:bss} we study the BSS population; in Section~\ref{sec:conclu} we discuss our results and summarize our conclusions.

\section{Dataset and Data Reduction}\label{sec:datared}

In this work we considered all the images from the Ultraviolet and Visible (UVIS) channel of the Wide Field Camera 3 (WFC3) and from the Wide Field Channel (WFC) of the Advanced Camera for Surveys (ACS) currently available in the \textit{HST} archive, covering the central regions of the cluster. The details of the observations are listed in Table~\ref{tab:1}.

\begin{table*}[t!]
\caption{List of \textit{HST} images of NGC~1261 used in this work.\label{tab:1}}
\centering
{
\begin{tabular}{llllll}
\hline\hline
Program ID & PI & Epoch & Camera & Filter & Exposures \\
 & & (yyyy/mm) & & & N $\times~t_{\mathrm{exp}}$ \\
\hline
GO-10775 & A. Sarajedini & 2006/03     & ACS/WFC & F606W & $1\times 40\,{\rm s}$\\
& &  &  &  & $ 6\times 350\,{\rm s}$\\
& & & & F814W & $1\times 40\,{\rm s}$\\
& &  &  &  &  $ 6\times 360\,{\rm s}$ \\
\hline
GO-13297 & G. Piotto & 2013/08 - 2014/06    & WFC3/UVIS & F275W & $2\times 834\,{\rm s}$\\
& &  &  &  & $ 4\times 855\,{\rm s}$ \\
& &  &  &  & $ 2\times 859\,{\rm s}$ \\
& &  &  &  & $ 2\times 918\,{\rm s}$ \\
& & & & F336W & $2\times 413\,{\rm s}$\\
& &  &  &  & $ 2\times 415\,{\rm s}$ \\
& &  &  &  & $ 1\times 419\,{\rm s}$ \\
& & & & F438W & $1\times 164\,{\rm s}$\\
& &  &  &  &$ 1\times 165\,{\rm s}$ \\
& &  &  &  &$ 1\times 167\,{\rm s}$  \\
& &  &  &  &$ 1\times 168\,{\rm s}$  \\
& &  &  &  &$ 1\times 170\,{\rm s}$  \\
\hline
GO-14235 & T. Sohn & 2017/03 & ACS/WFC & F606W & $1\times 45\,{\rm s}$\\
& &  &  &  & $ 7\times 607\,{\rm s}$ \\
& &  &  &  & $ 1\times 608\,{\rm s}$ \\

\hline
\end{tabular}}
\end{table*}

We performed a photometric reduction of the \texttt{\_flc} images, following the prescriptions given in \citet{bellini17a,bellini18}. We used \texttt{\_flc} images because the un-resampled pixel data for stellar profile fitting is preserved. These images have also been corrected to remove charge-transfer efficiency (CTE) defects, as described in \citet{andersonbedin10}. In the following, we briefly describe the reduction procedure, which consists of two main steps.

First, we performed a one-pass photometry, a single finding procedure without neighbour subtraction, to identify and measure all bright stars. We made use of spatially-variable point-spread-function (PSF) models tailored to each image starting from the publicly available library PSFs of the \textit{HST} detectors\footnote{Available at \url{http://www.stsci.edu/~jayander/STDPSFs/}} (e.g., \citealt{ak06}) by using a set of bright, relatively isolated and unsaturated stars in each exposure. These PSFs were used to measure stellar positions and fluxes in each exposure. Stellar positions were corrected for geometric distortion using the solutions provided in \citet{ak06,bellinibedin09,bellini11}.

Next, we run a multi-pass photometry using the software \texttt{KS2} (see \citealt{bellini17a} for details), which simultaneously performs the finding procedures on all the images and it is able to perform neighbor subtraction. This second step is necessary due to the high level of crowding in the central regions of GCs. \texttt{KS2} is able to simultaneously analyze multiple images taken with different filters, combining the results of the one-pass photometry transformed into a common reference frame.
As an astrometric reference system, we used the stellar positions in the \textit{Gaia} Data Release 2 (DR2) catalog (\citealt{gaia16,gaia18}).
We defined a common, pixel-based reference system, based on these Right Ascension (R.A.) and Declination (Dec.) positions, with X and Y increasing toward West and North, respectively, and with a pixel scale of 40~mas pixel$^{-1}$. We transformed the stellar positions as measured in each exposure onto this common reference frame system by means of six-parameter linear transformations. 
\texttt{KS2} performs the finding procedure through multiple iterations. For each iteration, it finds and subtracts stars that are present in a combination of filters chosen by the user, and then iteratively repeats this procedure for fainter stars that were not subtracted in the previous iteration, together with an improved fit and subtraction of already found stars. 
In this work, we applied an approach similar to that used in \citet{nardiello18}: we performed the first four finding iterations on the F606W and F814W exposures, while the fifth, sixth and seventh iterations were performed on the F438W, F336W and F275W images, respectively, to enhance the detection of blue stars that are too faint to be detected in the F606W and F814W exposures.
The instrumental magnitudes of each exposure were rescaled to that of the longest available exposure acquired with the same filter.

We calibrated our magnitudes to the \texttt{VEGAMAG} photometric system using the stars in common with the public catalog of NGC~1261 from the \textit{HST} UV Globular cluster Survey\footnote{Available at \url{http://groups.dfa.unipd.it/ESPG/treasury.php}.} (\citealt{piotto15,nardiello18}).

\begin{figure}[!t]
\centering
\includegraphics[width=0.97\textwidth]{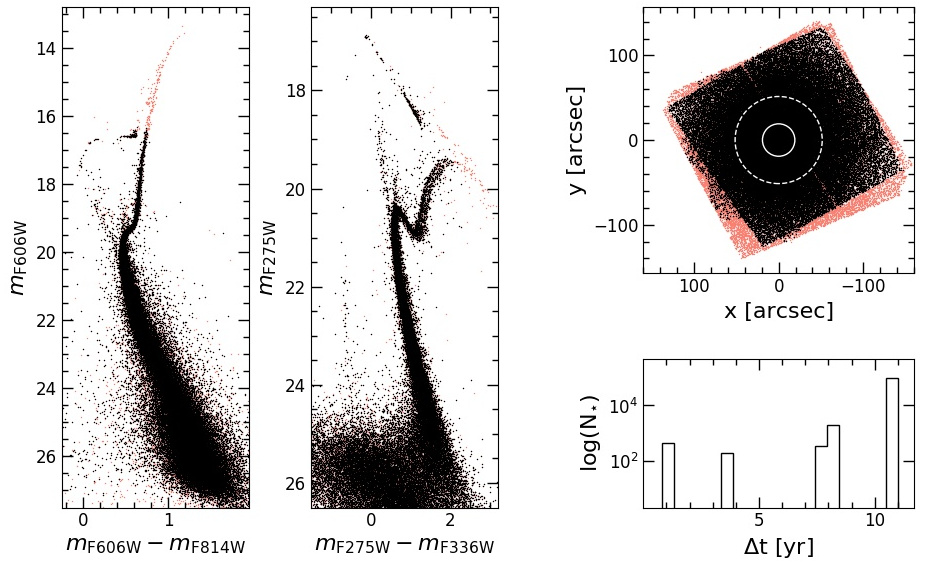}
\caption{Overview of the NGC~1261 catalog obtained in this work. In all panels, the red dots refer to the complete photometric catalog, while black dots are stars for which we also have a PM measurement. In the left and central panels we show the optical ($m_{\mathrm{F606W}}$ vs. $m_{\mathrm{F606W}}-m_{\mathrm{F814W}}$) and UV-blue ($m_{\mathrm{F275W}}$ vs. $m_{\mathrm{F275W}}-m_{\mathrm{F336W}}$) CMDs of NGC~1261. In these plots, we show all stars, regardless of their astro-photometric quality. The scattered cloud of points in the bottom left region of the UV-blue CMD is composed of stars that are measured in the F606W and/or F814W bands, but are too faint to be detected in the UV-blue bands. In spite of this, the software forces a fit, which results in an unreliable magnitude estimate that makes the star appear in this region of the CMD.
In the upper right panel we show the FOV covered by our dataset, with the center of the cluster (obtained as in Sect.~\ref{sec:struct}) placed in (0,0) and the projected positions of each star with respect to the center plotted along the x- and y-axes. North is up and East is to the left. In this panel we also show $r_{\rm{c}}$ and $r_{\rm{h}}$ as solid and dashed white circles, respectively, as obtained in Sect.~\ref{sec:struct}. In the lower right panel we show the number of objects (in a logarithmic scale) in our PM catalog, as a function of the temporal baseline of the observations used to measure them.}\label{fig:overview}
\end{figure}

We measured PMs using the technique developed by \citet{b14} and improved in \citet{bellini18} and \citet{libralato18}. We refer to those papers for a complete description of this procedure, which we briefly summarize in the following. This iterative procedure uses the stellar positions in each single exposure obtained from the multi-pass photometry. We transformed the \texttt{KS2}-based positions of each star in each image onto a common reference frame system. We fit these transformed positions as a function of the epoch with a least-square straight line. The slope of the straight line is a direct estimate of the PMs. The transformations to obtain master-frame positions are based on cluster member stars, therefore the PMs thus obtained are \textit{relative} to the cluster bulk motion. 
For this procedure we used all the catalogs available in our dataset (see Table~\ref{tab:1}), except for those in the F275W filter, which suffer from color-dependent systematic effects (\citealt{bellini11}), and therefore are not suitable to measure PMs.

In Figure~\ref{fig:overview}, we show an overview of the NGC~1261 catalog obtained in this work.
Black/red dots correspond to stars with/without a PM measurement. In the left and central panels we show the optical and UV-blue CMDs of NGC~1261. In the upper right panel we show the field of view (FOV) covered by our dataset, showing also the core and the half-mass radii ($r_{\rm{c}}$ and $r_{\rm{h}}$), obtained in Sect.~\ref{sec:struct}, as white circles. In the lower right panel we show the number of objects (in a log scale) in our PM catalog, as a function of the temporal baseline of the observations used to determine their PMs.

\section{Structural parameters}\label{sec:struct}

In order to derive the structural parameters of NGC~1261, we determined the center of gravity ($C_{\rm{grav}}$) and the star density profile. 

For the determination of $C_{\rm{grav}}$, we followed the iterative procedure described in \citet{montegriffo95}, and later used in \citet{lanzoni10, lanzoni19, miocchi13, raso17}. We measured $C_{\rm{grav}}$ as the average of the projected x and y positions of stars brighter than a given magnitude and within a given distance from the (first-guess) center. We chose an appropriate limiting magnitude around the TO level, to avoid incompleteness effects while maintaining a sufficient statistics. We assumed as a first-guess center the value of \citet{goldsbury10}, and iteratively recomputed the center, using each time the center obtained in the previous iteration, until the difference between two consecutive iterations was negligible. The resulting $C_{\rm{grav}}$ is located at R.A.$= 03^{\rm{h}}~12^{\rm{m}}~16.13^{\rm{s}}; \ \mathrm{Dec.}= -55^{\circ}~12^{\prime}~58.42^{\prime \prime}$, with uncertainties equal to: rms$_{\mathrm{R.A.}}=0.4^{\prime \prime}$ and rms$_{\mathrm{Dec.}}=0.2^{\prime \prime}$. 
The distance between the \citet{goldsbury10} center and the value of $C_{\rm{grav}}$ estimated in this work is equal to $\sim 0.7^{\prime \prime}$.

\begin{figure}[!t]
\centering
\includegraphics[width=\textwidth]{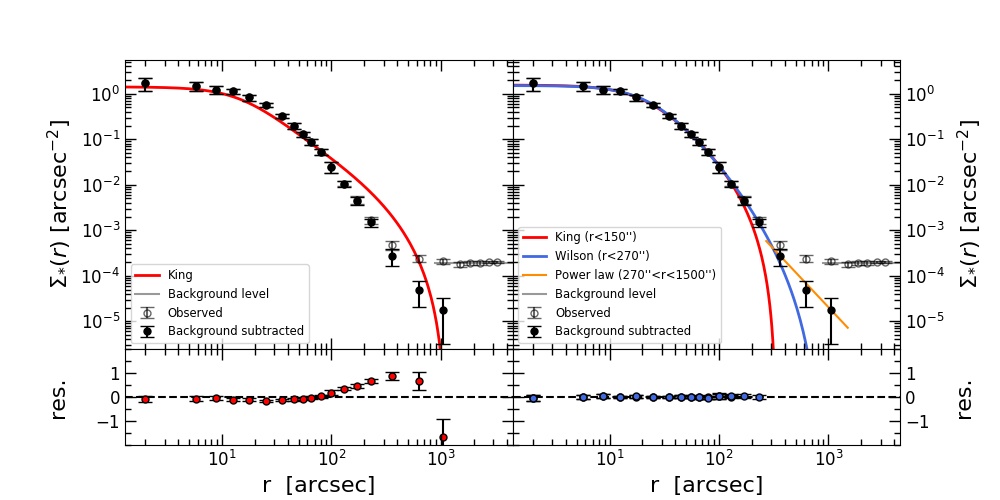}
\caption{Top panels: observed (open circles) density profile of NGC~1261. The leftmost 12 points have been obtained from the \textit{HST} data, while the remaining points have been obtained from \textit{Gaia} DR2 data. The assumed level of background, obtained as the average of the five rightmost points, is marked with a solid grey line. The background subtracted profile (filled circles), is obtained as the difference between the observed profile and the background level. In both top panels, the red line is the best fit King model obtained in this work using an MCMC approach (see text for details). In the top left panel, the King model fit has been performed along the entire extension of the background subtracted profile, while in the top right panel we only considered for the fit the points within $r=150^{\prime \prime}$. In the top right panel we also show the best fit Wilson profile (blue line), obtained considering only the points within $r=270^{\prime \prime}$. The yellow line is a power-law with slope $-2.6$, which approximately represents the behaviour of the outermost three points, where the profile cannot be fitted with neither the King nor the Wilson models, due to the presence of an outer envelope.
Bottom panels: residuals of the best-fit models, in a log scale (i.e., res = $\log \Sigma_{*,\mathrm{mod}} - \log \Sigma_{*,\mathrm{obs}}$). The residuals for the King model fit are shown as red circles, while the residuals for the Wilson model fit are shown in blue.}\label{fig:densprof}
\end{figure}

The available \textit{HST} data sample a region of $\sim 200^{\prime \prime} \times 200^{\prime \prime}$ roughly centered on the cluster center. This region is smaller than the tidal radius of NGC~1261 derived from the Harris catalog ($r_{\rm{t}}=304^{\prime \prime}$, \citealt{h96}). Therefore, in order to obtain a complete star density profile along the entire extension of the cluster, we combined the \textit{HST} data with \textit{Gaia} DR2 data. We constructed the star density profile as in \citet[see also \citealt{lanzoni10, lanzoni19}]{miocchi13}. We divided the sample into 23 concentric annuli\footnote{12 annuli for the \textit{HST} data, 13 for the \textit{Gaia} DR2 data, with 2 annuli in common between the two datasets. We used this region (70$^{\prime \prime}<$r$<110^{\prime \prime}$), where the \textit{Gaia} DR2 data do not suffer yet from incompleteness at the considered magnitude level, to normalize the \textit{Gaia} DR2 and \textit{HST} profiles.}, centered on $C_{\rm{grav}}$, considering only stars brighter than $m_{\rm{F606W}}=20$ (equivalent to $m_{\rm{G}}\sim 20$ for \textit{Gaia}), to avoid incompleteness effects and to properly combine the density profile obtained from \textit{HST} data with that determined from the \textit{Gaia} observations.
Each annulus was divided into four sectors, and the density in each sector was measured as the ratio between the number of stars within the sector and the area of the sector itself. The final density for each annulus corresponds to the mean of the four sector densities, and the density uncertainty is estimated from the variance among the sectors. The final, combined density profile is shown in the upper panels of Figure~\ref{fig:densprof}. The open circles represent the observed density profile, while the black circles represent the profile after background subtraction (measured as the average of the five rightmost points of the observed profile).

The density profile presented in Figure~\ref{fig:densprof} has been obtained using only stars brighter than the TO, i.e., stars with approximately the same mass. For this reason, the cluster structural parameters have been derived by fitting the density profile with single-mass, spherical and isotropic King models (\citealt{king66}).
We performed the fit using a Markov Chain Monte Carlo (MCMC) approach, using the \texttt{emcee} algorithm (\citealt{foremanmackey13}). We assumed uniform priors on the parameters of the fit, therefore the posterior probability distribution functions (PDFs) are proportional to the likelihood $\mathcal{L}=\exp(-\chi^2/2)$.

The result of this fit is shown in the left panel of Figure~\ref{fig:densprof}. While the agreement is satisfactory for $r < 100^{\prime \prime}$, the outer regions display significant deviations from the model. The disagreement between observations and best-fit model seems to be less severe in Figure E1 of \citet{baumgardthilker18}. However, this is because their density profile extends to just $\sim400^{\prime \prime}$ from the center, while the deviation from the King model distribution increases at large radii.  
This is consistent with the findings by \citet{kuzma18}, who detected the possible presence of an extended halo in the outer regions of this cluster.

\begin{table*}[t!]
\caption{Structural parameters obtained by fitting the observed number density profile with King (column 2) and Wilson (column 3) models, within a cluster centric distance $r=150^{\prime \prime}$ and $r=270^{\prime \prime}$, respectively.\label{tab:res_struct}}
\centering
{
\begin{tabular}{lll}
\hline\hline
Parameter & King model & Wilson model \\
\hline
\hline
$c$ & $1.19^{+0.07}_{-0.06}$ & $1.58^{+0.09}_{-0.07}$ \\
$r_{\rm{c}}$ & ($19.2^{+1.5}_{-1.4}$)$^{\prime \prime}$ & ($20.1\pm1.3$)$^{\prime \prime}$ \\
$r_{\rm{h}}$ & ($51.5^{+2.2}_{-2.0}$)$^{\prime \prime}$ & ($51.8^{+1.9}_{-1.8}$)$^{\prime \prime}$ \\
$r_{\rm{t}}$ & ($331.8^{+33.3}_{-27.4}$)$^{\prime \prime}$ & ($890.3^{+131.4}_{-103.7}$)$^{\prime \prime}$ \\
\hline
\end{tabular}}
\end{table*}

In order to further explore this topic, we limited the fit of the King models to the innermost region ($r<150^{\prime \prime}$) and considered also Wilson models (\citealt{wilson75}; see the right panels of Figure~\ref{fig:densprof}), which are more radially extended with respect to the King ones. 
As expected, Wilson models are able to reproduce the observed profile over a more extended radial portion (out to $270^{\prime \prime}$) with respect to the King family. However, an excess of stars is still observed at $r > 270^{\prime \prime}$, suggesting  the existence of an extended halo, in agreement with what found by \citet{kuzma18}. 
We highlight this feature with a power law with a slope of $-2.6$ (yellow line in the right panel of Figure~\ref{fig:densprof}). The power law we adopt is less steep than the one found by \citet{kuzma18}, however we do not intend here to analyze the characteristics of the outer envelope, but just to qualitatively highlight the presence of the envelope in the Figure. We refer to the aforementioned paper for an extensive analysis of the outer envelope of NGC~1261.

In Table~\ref{tab:res_struct} we list the best fit structural parameters (corresponding to the median of the PDF) and the uncertainties (68\% confidence interval) for the King and Wilson models. It is interesting to note that both the core radius and the half-mass radius agree between the two models at better than one arcsecond.
Our results are comparable, within the uncertainties, to the values obtained in \citet{kuzma18}. The only exception is the core radius, for which we obtain slightly smaller values for both the King and Wilson models. However, this is not concerning, since the difference is very small ($<2-3^{\prime \prime}$), and in general the agreement is good.

\section{Internal kinematics}\label{sec:kin}

\subsection{Proper motion selections}\label{ss:pmsel}

We selected stars for the kinematic analysis according to the following criteria: (i) at a given magnitude, they have a \texttt{QFIT} parameter (quality of the PSF fit, ranging from 0 to 1; larger values of \texttt{QFIT} indicates better quality of the PSF fit) larger than the 95-th percentile; (ii) their $o$ parameter (fraction of neighbour flux within the fitting radius before neighbour subtraction in the multi-pass photometry) is lower than 1; (iii) they were measured in at least 80\% of the images; (iv) at any given magnitude, the absolute value of their \texttt{RADXS} parameter (residual fractional source flux outside the fitting radius with respect to PSF predictions) is within $3\sigma$ from its median value; (v) their reduced $\chi^2$, obtained from the PM fit, is lower than 3; (vi) the fraction of measurements rejected in the PM fit procedure is less than 15\%; (vii) they are within 90$^{\prime \prime}$ from the cluster center (i.e. we excluded the edges of the FOV of the PM catalog, which is slightly smaller than the total FOV - see Figure~\ref{fig:overview}); (viii)  their PM error is smaller than 50\% the local (both in distance and in magnitude) intrinsic PM dispersion.

\begin{figure}[!t]
\centering
\includegraphics[width=\textwidth]{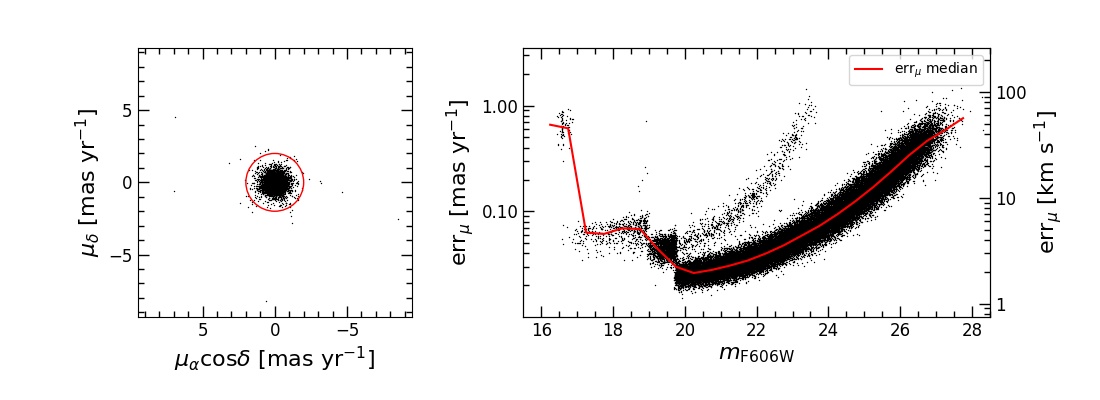}
\caption{Left panel: VPD of the stars satisfying criteria (i) to (vii), see text for details. The red circle corresponds to 2 mas yr$^{-1}$, marking the threshold adopted to reject non cluster members (note, however, the very low level of field contamination). Right panel: 1D PM error as a function of the F606W magnitude for the stars satisfying criteria (i) to (vii). The multiple sequences and discontinuities are due to the different temporal baselines and number of images used to compute the PMs in our project. The red line is the median value of err$_{\mu}$, measured in magnitude bins 0.5~mag wide.}\label{fig:vpdpmerr}
\end{figure}

The vector-point diagram (VPD, see the left panel of Figure~\ref{fig:vpdpmerr}) shows no prominent sign of field contamination, as expected from the Galactic latitude of this halo cluster (B=$-52.12^{\circ}$, \citealt{h96}). However, we applied a mild cut to exclude likely field stars, rejecting stars with a relative PM larger than 2 mas yr$^{-1}$ (i.e., more than 10 times the observed PM dispersion of the faintest stars detected; see the red circle in the left panel of Figure~\ref{fig:vpdpmerr}).

In the right panel of Figure~\ref{fig:vpdpmerr} we show the 1D PM error as a function of $m_{\mathrm{F606W}}$ for the stars satisfying criteria (i) to (vii). The red line corresponds to the median of the distribution, measured in bins 0.5~mag wide.
The different temporal baselines and number of images used to compute the PMs in our project cause the multiple sequences of PM errors clearly visible in the right panel of Figure~\ref{fig:vpdpmerr}.

The observed velocity dispersion obtained from PMs is the combination of the true velocity dispersion of the cluster and the scatter due to PM errors, which must be subtracted in quadrature in order to obtain the true dispersion. Thanks to state-of-the-art reduction techniques, as the ones used in this work and in \citet{b14, bellini15, bellini18, libralato18, libralato19}, we are able to keep the PM errors as low as possible. 
However, for such a distant GC, stellar PMs are small compared to the PM uncertainty and in some cases we were unable to derive meaningful velocity dispersions, even after the subtraction of the PM uncertainty.
Hence, for the kinematic analysis presented in Section~\ref{sec:vdispprof}, we restricted our sample to the magnitude range $19.75 < m_{\mathrm{F606W}} < 20.75$. This condition has the double benefit of allowing us to study the kinematics of stars of comparable masses, while using the stars with PM errors as low as possible.

\subsection{Velocity dispersion profiles}\label{sec:vdispprof}

In this section we present and discuss the velocity dispersion profiles of NGC~1261 over its entire radial extension, derived by combining the kinematical information in the plane of the sky, obtained from PMs, for the cluster central regions with LOS velocities of individual stars, measured in the external regions from the MIKiS survey (\citealt{ferraro18mikis}).

The radial and tangential velocity dispersion profiles in the plane of the sky, obtained from our PMs, were measured maximizing the following likelihood in different radial bins:

\begin{equation}\label{eq:like}
\ln \mathcal{L} = -\frac{1}{2} \sum_{n} \Bigl[\frac{v_{\mathrm{rad},n}^2}{\sigma_{\mathrm{rad}}^2+\epsilon_{\mathrm{rad},n}^2} + \ln(\sigma_{\mathrm{rad}}^2 + \epsilon_{\mathrm{rad},n}^2) + \frac{v_{\mathrm{tan},n}^2}{\sigma_{\mathrm{tan}}^2+\epsilon_{\mathrm{tan},n}^2} + \ln(\sigma_{\mathrm{tan}}^2 + \epsilon_{\mathrm{tan},n}^2) \Bigr]    
\end{equation}
where $v_{\mathrm{rad},n}$ and $v_{\mathrm{tan},n}$ are the radial and tangential components of the velocity of each star in the considered bin, $\epsilon_{\mathrm{rad},n}$ and $\epsilon_{\mathrm{tan},n}$ are the observed errors on the radial and tangential velocity components, and $\sigma_{\mathrm{rad}}$ and $\sigma_{\mathrm{tan}}$ are the radial and tangential velocity dispersions, respectively.
We used again a MCMC approach, based on the \texttt{emcee} algorithm (\citealt{foremanmackey13}), obtaining the posterior probability distribution functions (PDFs) for $\sigma_{\mathrm{rad}}$ and $\sigma_{\mathrm{tan}}$. The best-fit values correspond to the PDF median, while the resulting uncertainties correspond to the 68\% confidence interval. 

\begin{figure}[!t]
\centering
\includegraphics[width=0.9\textwidth]{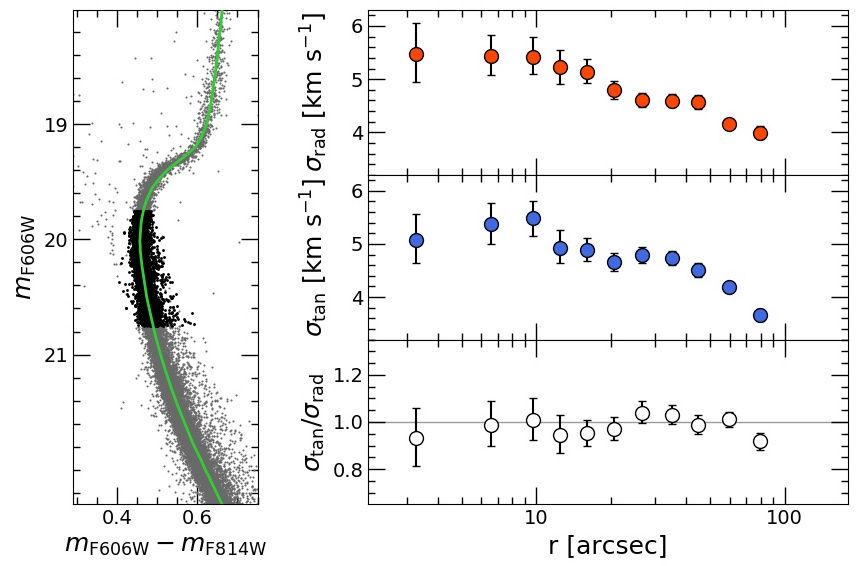}
\caption{Left panel: CMD of NGC~1261 ($m_{\mathrm{F606W}}$ vs. $m_{\mathrm{F606W}}-m_{\mathrm{F814W}}$), using the quality-selected PM catalog (grey dots). The black points show the sub-sample of MS-TO stars selected to compute the velocity dispersion profiles. The green line is a PARSEC isochrone (\citealt{marigo17}), with an age of 11.5~Gyr and [Fe/H]=$-$1.35 (age and metallicity taken from \citealt{dotter10}). Right panels, from top to bottom: radial and tangential velocity dispersion profiles, and anisotropy profile (plotted respectively as red, blue and white circles).}\label{fig:dispprofs}
\end{figure}

In order to allow a comparison between the velocity dispersion profiles presented here with previously published profiles, mainly obtained from radial velocities from spectroscopy (therefore measured for a sample of bright stars, with a mass comparable to the TO mass), we must only consider the kinematics of stars of similar mass, because of the effects of energy equipartition. The restriction of our sample to the stars in the $m_{\mathrm{F606W}}$ range between 19.75 and 20.75, discussed in the previous section and made to select only stars with the lowest PM errors, serves also this purpose. We excluded stars brighter than $m_{\mathrm{F606W}}$=19.75 because, as it can be seen in the right panel of Figure~\ref{fig:vpdpmerr}, at brighter magnitudes the PM errors increase significantly, due to the smaller number of images available for the PM computation (the longest exposures are saturated above this magnitude value). We chose $m_{\mathrm{F606W}}$=20.75 as our faint limit to include in our sample only stars with masses comparable to the TO mass.
The masses in this magnitude interval, estimated from a PARSEC isochrone (\citealt{marigo17}) with an age of 11.5~Gyr and [Fe/H]=$-$1.35 (green line in the left panel of Figure~\ref{fig:dispprofs}; age and metallicity are taken from \citealt{dotter10}), range from $\sim$0.75~M$_{\odot}$ to $\sim$0.8~M$_{\odot}$, while bright stars (e.g., Red Giant Branch stars) in NGC~1261, according to the same isochrone, have masses up to $\sim$0.82~M$_{\odot}$. Therefore, even if velocity dispersion decreases when mass increases due to energy equipartition ($\sigma \propto m^{-\eta}$), in such a small mass range the velocity dispersion difference is expected to be negligible. Even in the extreme case of full energy equipartition ($\eta$=0.5), which is not expected to be reached (e.g., \citealt{trentivandermarel13, bianchini16}), the difference in velocity dispersion for the considered mass range is at most a few tenths of km s$^{-1}$. We also used a mild color cut ($0.4 < m_{\mathrm{F606W}}-m_{\mathrm{F814W}} < 0.6$) to exclude the most obvious outliers. This mild color cut does not exclude binaries, which have a different kinematics with respect to single stars. However, the binary fraction of NGC~1261 is lower than 5\% (\citealt{milone12}), therefore their contribution should be negligible.

Figure~\ref{fig:dispprofs} summarizes our results on the kinematics of MS-TO stars in NGC~1261. In the left panel we show as grey points the CMD of stars in NGC~1261 that have a PM measurement and pass our quality selections (see Section~\ref{ss:pmsel}). The MS-TO stars selected for the kinematical analysis (5410 stars) are shown as black points. 
In the right panels of Figure~\ref{fig:dispprofs} we show the velocity dispersion profiles\footnote{PMs in mas yr$^{-1}$ have been converted into km s$^{-1}$, through the equation $\sigma_{\mathrm{km~s^{-1}}} = 4.74~D_{\mathrm{kpc}}~\sigma_{\mathrm{mas~yr^{-1}}}$ (e.g., \citealt{vandeven06}).} obtained in this work. 
From top to bottom, we show the radial velocity dispersion profile ($\sigma_{\mathrm{rad}}$, red circles), the tangential velocity dispersion profile ($\sigma_{\mathrm{tan}}$, blue circles), and the anisotropy profile ($\sigma_{\mathrm{tan}}/\sigma_{\mathrm{rad}}$, white circles).

The central region of NGC~1261 appears to be isotropic ($\sigma_{\mathrm{tan}}/\sigma_{\mathrm{rad}}\sim 1$); the only point in the anisotropy profile that deviates significantly from isotropy is the most external one, at a distance from the center of about 80$^{\prime \prime}$, where the cluster appears to be mildly radially anisotropic. 
Although this trend is consistent with both simulations and observations, which find that some GCs are isotropic in the central regions and radially anisotropic at larger radii (see e.g., \citealt{vesperini14,tiongco16} for simulations and \citealt{b14, bellini17d,watkins15a} for observational results), this concerns only one point at the edge of the FOV sampled by our observations, corresponding to $\sim2~r_{\mathrm{h}}$ from the center. Hence, from now on we make the conservative assumption of an overall isotropy for NGC~1261.

\begin{figure}[!t]
\centering
\includegraphics[width=0.8\textwidth]{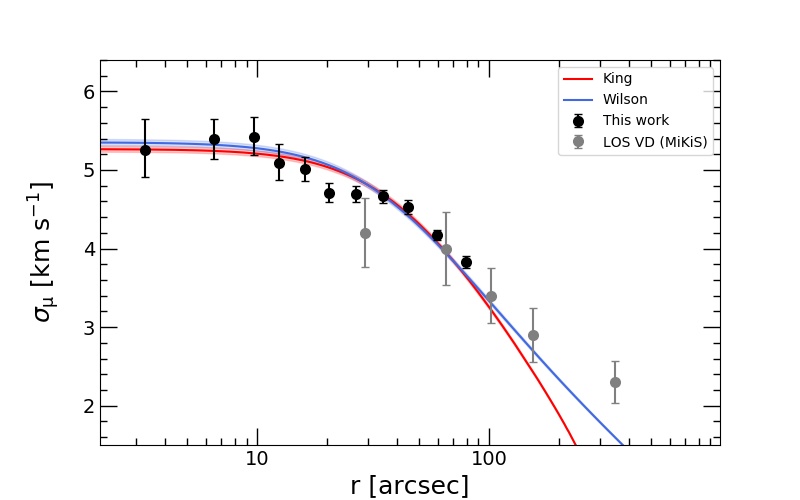}
\caption{Global velocity dispersion profile of NGC~1261. The measurements in the plane of the sky, obtained from \textit{HST} PMs are plotted as black circles, while the grey circles correspond to the LOS velocity dispersion profile from \citet{ferraro18mikis}. The best-fit King and Wilson profiles are shown as red and blue lines, respectively. The shaded areas around the two best-fit profiles represent the uncertainty on the central velocity dispersion (see text for details on the fitting procedure).}\label{fig:dispproffit}
\end{figure}

Given the overall isotropy of the sampled region, we determined the global velocity dispersion profile in the plane of the sky ($\sigma_{\mu}$), shown in Figure~\ref{fig:dispproffit} (black circles). This has been obtained with the same procedure described above, but imposing $\sigma_{\mu}=\sigma_{\mathrm{rad}}=\sigma_{\mathrm{tan}}$ in the likelihood. We combined the global velocity dispersion profile in the plane of the sky with the LOS velocity dispersion profile from \citet{ferraro18mikis}. As previously discussed, the two velocity dispersion profiles can be compared because they are obtained from stars that are roughly in the same mass range. As shown in the Figure, the two profiles nicely connect in the radial region in common, with no need for arbitrary vertical shifts.

We then fitted the combined (PMs+LOS) velocity profile with King and Wilson models. We used a MCMC fitting procedure equivalent to the one described previously.  We kept the structural parameters obtained in Sect.~\ref{sec:struct} fixed and we adopted only the central velocity dispersion, $\sigma_0$, as a free parameter\footnote{We want to point out that the fit can be performed also without fixing the structural parameters to the values obtained from the density profile fit. However, the determination of the structural parameters from the velocity dispersion profile is more uncertain than using the density profile, due to the smaller number of points and higher probability of systematics.}.
For consistency with the density profile fitting procedure (see Section~\ref{sec:struct}), we excluded the regions beyond $r=150^{\prime \prime}$ and $r=270^{\prime \prime}$ for the King and Wilson models, respectively. The observed velocity dispersion profile is reasonably well reproduced by the adopted models (red and blue lines in Figure~\ref{fig:dispproffit}), with central dispersion values of $\sigma_{0,\mathrm{K}}=5.27\pm0.04$ km s$^{-1}$ and $\sigma_{0,\mathrm{W}}=5.35\pm0.04$ km s$^{-1}$, respectively.
These values are comparable to previous estimates based only on radial velocities (thus limited to more external regions of the cluster), e.g., \citet{baumgardthilker18} obtain $\sigma_0$=5.6~km s$^{-1}$; \citet{ferraro18mikis} obtain $\sigma_0$=5.5$\pm$0.4~km s$^{-1}$. 

The central velocity dispersion measurements can be used to estimate the total mass of NGC~1261, by properly rescaling and integrating the best fit King and Wilson density profiles. We obtained $M=(1.14^{+0.14}_{-0.12}) \times 10^5 \mathrm{M_{\odot}}$ from the King model and $M=(1.13^{+0.12}_{-0.10}) \times 10^5 \mathrm{M_{\odot}}$ from the Wilson model. 
Although these values formally underestimate the total mass of the system (since they have been derived from the King/Wilson models), the contribution of the stars populating the halo surrounding the cluster is likely negligible. Indeed, the two estimates are very similar (in spite of a larger extension of the Wilson model, with respect to the King one), and also consistent with the value quoted in \citet[$M=1.30 \times 10^5 \mathrm{M_{\odot}}$]{ferraro18mikis}.

\subsection{Absolute proper motion}\label{sec:abspm}

As already mentioned, the PMs presented here are \textit{relative} to the bulk motion of the cluster. Its absolute motion is consequently transferred, with the opposite sign, to background objects that, in reality, are essentially not moving with respect to it (e.g., distant galaxies; see, e.g., \citealt{massari13, cadelano17, libralato18abspm}).

\begin{figure}[!t]
\centering
\includegraphics[width=\textwidth]{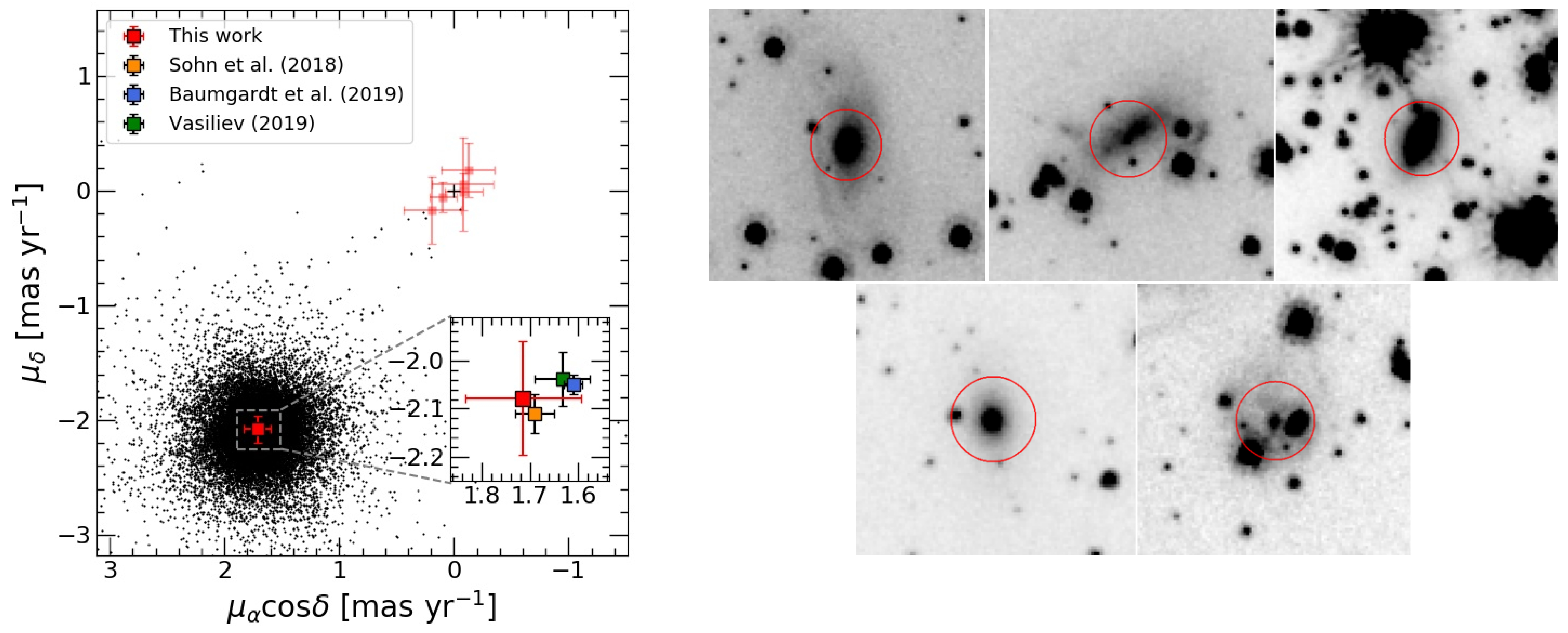}
\caption{Left panel: VPD of the absolute proper motions. The light red squares are the PMs of 5 background galaxies found by visually inspecting the images. The cluster absolute proper motion is marked as a dark red square. In the inset we zoom in around the cluster absolute proper motion and we show as yellow, blue and green squares the absolute motion from \citet{sohn18}, \citet{baumgardt19} and \citet{vasiliev19}, respectively. Right panel: images in the F814W band of the 5 galaxies used to measure the absolute proper motion of NGC~1261.}\label{fig:abspm}
\end{figure}

We visually inspected the images in our dataset, looking for background galaxies with sharp nuclei, in order to be able to obtain a meaningful photometry and astrometry with a stellar PSF. We found 5 background galaxies with reliable PMs. Their finding charts are shown in the right panel of Figure~\ref{fig:abspm}. These galaxies are distributed within the entire field of view, which minimizes the effects of cluster rotation, if present. We were not able to directly study cluster rotation from our PMs, because, during the PM computing procedure, we use six-parameter linear transformation to transform each star position into a given reference frame. These transformation also solve for rotation, therefore we cannot study rotation from cluster-member PMs. Any possible sign of rotation is transferred, with the opposite sign, to non-cluster objects in the field. Therefore, we can in principle use this property to study cluster rotation (see, e.g., \citealt{massari13, bellini17d, libralato18}), however no suitable field object can be found in this particular case to properly perform a rotation analysis.

The absolute PMs of the 5 galaxies that we found are shown in the left panel of Figure~\ref{fig:abspm} as light red squares. The absolute proper motion of NGC~1261, shown as a dark red square in Figure~\ref{fig:abspm}, that we consequently obtain is:

\begin{equation}
(\mu_{\alpha} \cos \delta, \mu_{\delta})_{\mathrm{NGC~1261}} = (1.71 \pm 0.12, -2.08 \pm 0.12) \ \mathrm{mas~yr^{-1}}
\end{equation}

This result is consistent, within the errors, with the results of \citet{sohn18}, obtained from \textit{HST} data and a data reduction technique specifically designed to well measure the position of galaxies, and \citet{baumgardt19} and \citet{vasiliev19}, obtained from \textit{Gaia} DR2 data. 
Their results are shown in the inset of Figure~\ref{fig:abspm} as yellow, blue and green squares, respectively. \\ \\

\section{Blue Straggler Stars}\label{sec:bss}

\subsection{BSS Selection}\label{ss:bsssel}

As already discussed in \citet[see also \citealt{ferraro18treasury}]{raso17}, a purely UV CMD (e.g., $m_{\mathrm{F275W}}$ vs. $m_{\mathrm{F275W}}-m_{\mathrm{F336W}}$) is the ideal diagram to select BSSs. In this diagram, BSSs define an almost vertical sequence, clearly distinguishable from other evolutionary sequences. Also, at these wavelengths, the cooler stellar populations (e.g., Red Giant Branch, Asymptotic Giant Branch), which can severely limit our capability of detecting complete samples of BSSs (see Figures 4, 5 and 6 in \citealt{raso17}), are significantly fainter, while BSSs are among the brightest objects.

\begin{figure}[!t]
\centering
\includegraphics[width=0.85\textwidth]{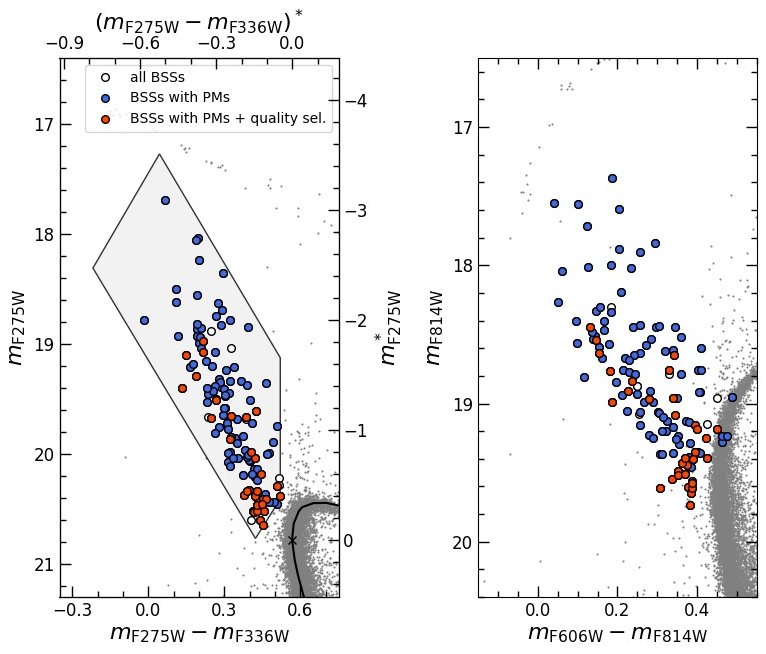}
\caption{Left panel: UV CMD of the BSS region of NGC~1261. The shaded box with black contours is the BSS selection box (see text for details). The axis of the n-CMD (i.e., a CMD with color and magnitude rigidly shifted to locate the TO at (0,0)) are reported on the right and top axis. The isochrone and the relative TO point are shown as a black curve and a black cross, respectively. All the stars that fall within the selection box in this CMD are selected as BSSs (black empty circles). We show as blue solid circles all the BSSs that have a measured PM, and in red the BSSs with a measured PM that also pass the selection criteria described in Section~\ref{ss:pmsel}. Right panel: optical $m_{\mathrm{F814W}}$ vs. $m_{\mathrm{F606W}}-m_{\mathrm{F814W}}$ CMD, with the UV-selected BSSs shown with the same color code adopted in the left panel.}\label{fig:bsssel}
\end{figure}

Therefore, we used the equations reported in Section 3 of \citet{raso17} to define the BSS selection box in the normalized CMD (n-CMD)\footnote{The n-CMD is a CMD where the stellar magnitudes and colors have been rigidly shifted so that the MS-TO is at $m^*_{\mathrm{F275W}}=0$ and $(m_{\mathrm{F275W}}-m_{\mathrm{F336W}})^*=0$. This is useful when comparing GCs of different metallicity, as in \citet{raso17} and \citet{ferraro18treasury}, because the BSS sequence is always located in the same region of the n-CMD. For consistency, here we adopt the same procedure, even if we are analyzing a single GC.} of NGC~1261. In the left panel of Figure~\ref{fig:bsssel} we show the selection box (black thick line and shaded area) and the selected BSSs in the UV CMD (large circles). We also highlight the BSSs that have a measured PM (blue circles) and BSSs with PMs that pass our selection criteria for kinematics, discussed in Section~\ref{ss:pmsel} (red circles). In total, we selected 132 BSSs; 122 BSSs have a PM measurement, 32 of which pass the kinematic selection criteria.  In the right panel of the Figure we show the selected BSSs in an optical CMD, with the same color code as in the left panel. As already discussed in \citet{raso17}, the BSS selection performed in the UV CMD allows us to also safely select very faint BSSs, which in an optical CMD are extremely close to the MS-TO or even overlapping the MS-TO/Sub-Giant Branch cluster population.

\begin{figure}[!t]
\centering
\includegraphics[width=\textwidth]{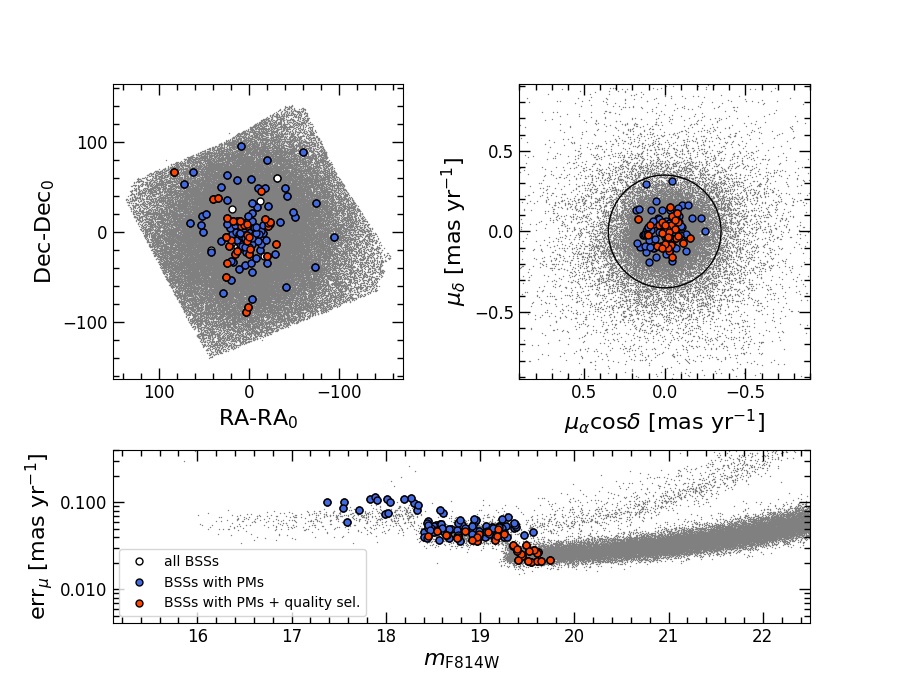}
\caption{Top left panel: map of the FOV analyzed in this work. We report in grey all the stars from the photometric catalog, and the selected BSSs as large circles. Here and in the other panels, we use for BSSs the same color code adopted in Figure~\ref{fig:bsssel}, i.e.: empty circles for the complete sample of BSSs, blue circles for BSSs with a PM measurement, red circles for BSS with a PM measurement that pass the kinematic selection criteria described in Section~\ref{ss:pmsel}. As expected, BSSs are placed preferentially in the central region of the cluster. 
Top right panel: a zoomed version of the VPD shown in the left panel of Figure~\ref{fig:vpdpmerr}. We show in grey the stars satisfying criteria (i) to (vii) (see Section~\ref{ss:pmsel}). The black circle corresponds to three times the standard deviation of the PM distribution for stars in a magnitude range comparable to BSSs ($17<m_{\mathrm{F814W}}<20$). Although this value represents a reasonable threshold to separate member BSSs from field contaminants, all BSSs (blue and red circles) fall within this threshold value.
Bottom panel: 1D PM error with respect to the F814W magnitude. Clearly, BSS PM errors become larger for increasing stellar luminosity. This is due to the effect of saturation, occurring at different magnitudes for different datasets and thus reducing the total time baseline and/or the number of available exposures that can be used to compute PMs.}\label{fig:mappm}
\end{figure}

In the top left panel of Figure~\ref{fig:mappm} we show the map of the FOV analyzed in this work, with selected BSSs highlighted with the same color code as in Figure~\ref{fig:bsssel}. As expected, BSSs are found to preferentially populate the central regions of the cluster.
In the top right panel of Figure~\ref{fig:mappm}, we show a zoomed version of the VPD plotted in Figure~\ref{fig:vpdpmerr}. 
The black circle corresponds to three times the standard deviation of the PM distribution of all stars with magnitude comparable to that of BSSs (roughly, $17<m_{\mathrm{F814W}}<20$). From this plot, it appears clear that the selected BSSs are members of NGC~1261. 
In the bottom panel of Figure~\ref{fig:mappm} we show the 1D PM error as a function of the F814W magnitude. 
BSSs are located exactly where the PM errors worsen. This is because the brightest BSSs are saturated in most of the deep optical exposures, and for this reason they have been measured only in a sub-sample of the available images. This reduced the temporal baseline of the PM measurements and the quality of the derived PMs. Only the faintest BSSs have high quality PMs; therefore, when using the BSSs sample selected with the kinematic quality criteria (see Section~\ref{ss:bsskin}; red circles in Figure~\ref{fig:mappm}) we must keep in mind that we are preferentially selecting faint BSSs, i.e., low mass BSSs (see \citealt{raso19}).

\subsection{BSS radial distribution}\label{ss:apiu}

\begin{figure}[!t]
\centering
\includegraphics[width=\textwidth]{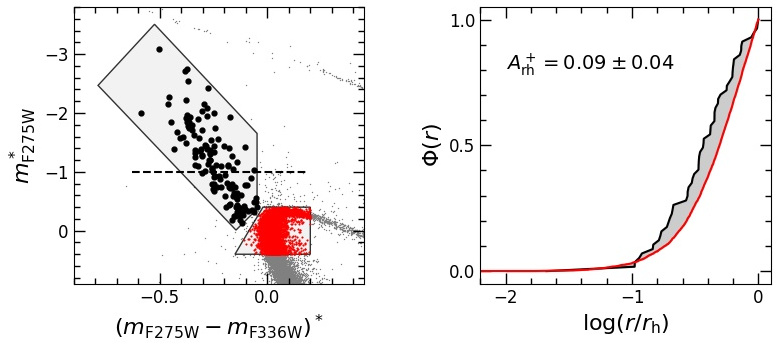}
\caption{Left panel: n-CMD of NGC~1261. We show as black circles the selected BSSs. The black dashed line corresponds to the magnitude threshold adopted to measure $A^+_{\mathrm{rh}}$ (only BSSs brighter than this magnitude have been used). The red dots are the MS-TO stars used as reference population. Right panel: cumulative radial distribution of BSSs (black line) and MS-TO reference population (red line). The resulting value of $A^+_{\mathrm{rh}}$ is labeled within the panel.}\label{fig:apiu}
\end{figure}

As already mentioned, BSSs are a population of heavy stars (with an average mass of $\sim 1.2 \mathrm{M_{\odot}}$; see \citealt{shara97, gilliland98, ferraro06, fiorentino14, baldwin16, raso19}), while the average stellar mass in GCs is significantly smaller, $\langle m\rangle \sim 0.4 \mathrm{M_\odot}$. Therefore, BSSs are subject to dynamical friction, which progressively makes them sink toward the cluster center. For this reason, the BSS radial distribution is a fundamental tool to measure the dynamical evolution of GCs. In particular, \citet{alessandrini16} and \citet{lanzoni16} introduced the indicator $A^+$, able to quantify the level of dynamical evolution of a cluster. This indicator is defined as the area enclosed between the cumulative radial distribution of BSSs and the one of a reference population of lighter stars. The distance from the center is expressed in logarithmic units in order to maximize the sensitivity of the indicator in the central regions, where dynamical friction accumulates most of the BSSs. 
The value of $A^+$ depends on the adopted distance from the cluster center. Thus, in order to perform meaningful comparisons among different  clusters, \citet{lanzoni16} measured this parameter within one half-mass radius ($A^+_{\mathrm{rh}}$). Recently, \citet{ferraro18treasury} investigated a sample of 48 Galactic GCs and found a strong correlation between $A^+_{\mathrm{rh}}$ and the number of central relaxation times experienced by the host star cluster.
This result confirms the efficiency of $A^+$ as a tool to measure the dynamical evolution of GCs. Furthermore, \citet{ferraro19}, explored a sample of 5 old Large Magellanic Cloud GCs, probing the presence of a correlation between $A^+_{\mathrm{rh}}$ and the number of central relaxation times also in stellar systems outside the Milky Way.

NGC~1261 is included in the sample discussed by \citet{ferraro18treasury}, who found $A^+_{\mathrm{rh}} = 0.10\pm0.02$ for this cluster. However, here we re-determine its value using the new center and half-mass radius obtained in this work. Both quantities can have a significant impact on the determination of $A^+_{\mathrm{rh}}$, and, in particular, the half-mass radius adopted by \citet[$r_{\mathrm{h}}=40.8^{\prime \prime}$, from \citealt{h96}]{ferraro18treasury} is significantly smaller than the value obtained here ($r_{\mathrm{h}}\simeq52^{\prime \prime}$, see Table~\ref{tab:res_struct}).

For consistency with previous estimates, to measure $A^+_{\mathrm{rh}}$ we used BSSs brighter than $m^*_{\mathrm{F275W}}=-1$ in the n-CMD (see the dashed horizontal line in the left panel of Figure~\ref{fig:apiu}). This way, we select only the heaviest BSSs, for which the effect of dynamical friction is stronger, and we exclude the part of the BSS sequence where there could be a contamination from MS-TO stars or blends. As reference population, we used MS-TO stars, highlighted in red in the left panel of Figure~\ref{fig:apiu}. In the right panel we show the cumulative radial distributions of BSSs (black line) and MS-TO reference stars (red line). The shaded grey area between the two curves corresponds to the measured value of $A^+_{\mathrm{rh}}$, equal to $0.09\pm0.04$. The errors on $A^+_{\mathrm{rh}}$ have been estimated by combining the uncertainties due to small-number statistics, and the uncertainties on the half-mass radius and the position of the center. The former, which turned out to be the primary source of uncertainty, has been estimated with a jackknife bootstraping tecnique (\citealt{lupton93}), as in \citet{ferraro18treasury}. The latter has been estimated by repeating the measure of $A^+_{\mathrm{rh}}$ 1000 times, each time randomly extracting the values of the center and half-mass radius from a Gaussian distribution with mean and dispersion equal to the estimates obtained in this work.

The derived value of $A^+_{\mathrm{rh}}$ is in very good agreement with that quoted by \citet{ferraro18treasury}, and indicates that NGC~1261 experienced a moderate level of BSS central sedimentation, thus suggesting a moderate level of dynamical evolution. In particular, by using Eq. (2) in \citet{ferraro18treasury}, we obtain $N_{\mathrm{relax}}=t_{\mathrm{age}}/t_{\mathrm{rc}} = 18^{+12}_{-7}$, corresponding to a relatively large central relaxation time: $t_{\mathrm{rc}}=0.65^{+0.41}_{-0.26}$ Gyr. 
For comparison, the most dynamically evolved clusters in the sample analyzed by \citet{ferraro18treasury}, display values of $A^+_{\mathrm{rh}}$ as large as 0.4, with $N_{\mathrm{relax}}>300$. These considerations suggest that NGC~1261 is not in an advanced stage of dynamical evolution (see also \citealt{dalessandro19apiu}).

\subsection{A double sequence in NGC~1261?}

The BSS population of NGC~1261 has been studied by \citet{simunovic14}, from the analysis of a subset of the dataset used in this work (the F606W and F814W images from GO-10775, and the F336W images from GO-13297 - see Table~\ref{tab:1} for details). They found that BSSs in this cluster define at least two sequences. In Galactic GCs, this feature has been observed in M30 (\citealt{ferraro09}) and NGC~362 (\citealt{dalessandro13, libralato18}), and recently also in M15 (\citealt{beccari19}). 
\begin{figure}[!t]
\centering
\includegraphics[width=0.8\textwidth]{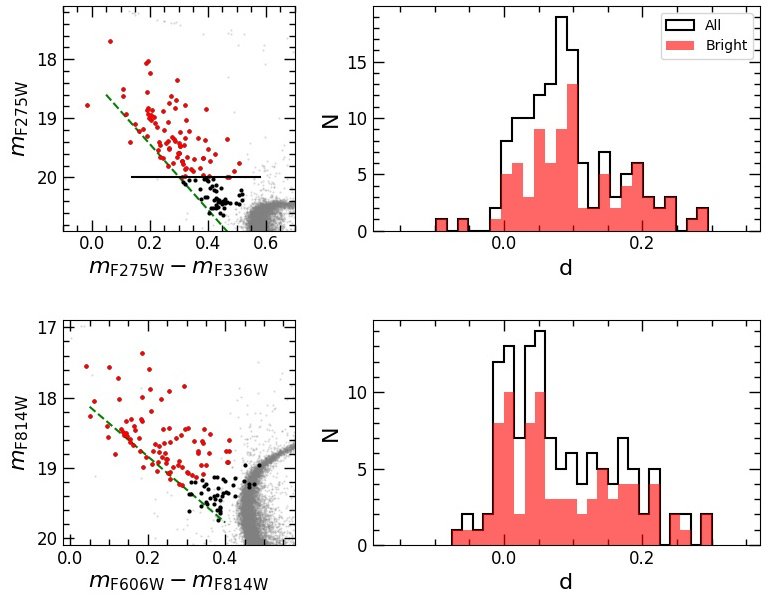}
\caption{Top left panel: UV CMD of the BSS region of NGC~1261. The selected BSSs are highlighted as red and black circles. The green dashed line roughly corresponds to the blue edge of the bulk of the BSS sequence, excluding a few scattered, blue objects. We highlight in red the BSS brighter than $m_{\rm F275W}=20$. Top right panel: histogram of the distances of BSSs from the green dashed line in the top left panel. The black histogram shows the distribution of all the BSSs in our sample, while the red histogram shows the distribution of the BSSs brighter than $m_{\rm F275W}=20$. Bottom left panel: optical CMD of the BSS region of NGC~1261. The same selections and color code as in the top left panel are adopted. Bottom right panel: as in the top right panel, but for the distances of BSSs from the green line in the optical CMD.}\label{fig:doubleseq}

\includegraphics[width=0.45\textwidth]{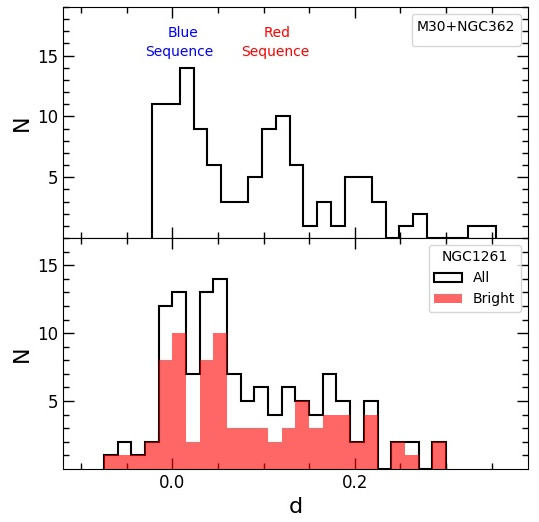}
\caption{Histograms of the distances of BSSs in M30+NGC~362 (top panel) and in NGC~1261 (bottom panel). The red histogram in the bottom panel shows the distribution of BSSs brighter than $m_{\mathrm{F275W}}=20$.}\label{fig:doubleseqconfro}
\end{figure}

The main characteristic of this feature is the presence of a narrow blue BSS sequence, separated by a clear gap from a broader red BSS sequence. 
As first suggested by \citet{ferraro09}, the narrowness of the blue sequence and its impressive agreement with the locus of BSS collisional isochrones (\citealt{sills09}) point towards a simultaneous and collisional origin of these BSSs, possibly related to the cluster core collapse (CC), with the possibility of even dating the epoch of such event (see also \citealt{portegies19}).

We checked our photometry for this feature. In Figure~\ref{fig:doubleseq} we show both the UV CMD used for the BSS selection (top left panel, see Section~\ref{ss:bsssel}), and the optical CMD (bottom left panel). The optical CMD is remarkably similar to that published by \citet{simunovic14}, and a vague hint of a bimodal distribution can be distinguished by eye, especially if the faintest portion of the BSS sequence is not considered. In order to quantify this impression, in both diagrams, we draw a straight line roughly corresponding to the blue edge of the BSS sequence (green, dashed lines in Figure~\ref{fig:doubleseq}), excluding just a few scattered, blue objects, and we compute the distance of each BSS from those lines (perpendicularly to the line itself). The distributions of the resulting BSS distances is plotted in the right panels of Figure~\ref{fig:doubleseq}, both for the complete sample of BSSs (black histogram) and for a selected sample, excluding faint BSSs ($m_{\mathrm{F275W}}<20$, red histogram). As can be seen, no evidence of bimodal distribution is visible in the distribution of BSS distances computed in the UV CMD. Instead, the optical CMD shows a double peak, very similar to that discussed by \citet{simunovic14}. 

This is found, however, only if quite narrow bins are adopted, while using wider bins reduces or completely erases the feature. Moreover, the bin width ($\sim0.015$ mag) necessary to highlight two distinct peaks in the distribution is comparable to the average photometric error of the BSS population ($\sim0.01$ mag). This clearly casts doubts about the real existence of a double BSS sequence in NGC~1261. We also highlight that the distance between the two BSS peaks detected in M30 and NGC~362 ($\sim0.1$ mag) is several times larger than in NGC~1261. This is clearly appreciable in Figure~\ref{fig:doubleseqconfro}, where the combined distribution of the BSS distances for M30 and NGC~362 (upper panel) is compared to that obtained in NGC~1261 (lower panel). Indeed, the bimodal feature detected in NGC 1261 seems to be morphologically different from those discovered in M30 and NGC~362: the two peaks identified in NGC~1261 are essentially located within the ``collisional peak'' of the other two clusters. On the other hand, the measure of the BSS sedimentation level, obtained in Section~\ref{ss:apiu}, suggests that NGC~1261 is still far from its CC phase, quite differently from M30 and NGC~362, which are both in very advanced stages of dynamical evolution. This comparison thus poses questions about the similarity of these features: even if confirmed by future observations with higher photometric accuracy, the double BSS sequence in NGC~1261 might have a different origin from that suggested for M30 and NGC~362 (\citealt{ferraro09,portegies19}).

\subsection{BSS kinematics}\label{ss:bsskin}

GCs tend to evolve towards a state of energy equipartition, i.e., energy exchanges among stars of different masses redistribute kinetic energy. As a consequence, lighter stars increase their velocity, while heavier stars (like BSSs) tend to slow down. Therefore, it is interesting to compare the velocity dispersion profile of ``normal'' stars (e.g., MS-TO stars, which have a mass of $\sim 0.8~\mathrm{M_{\odot}}$; see Section~\ref{sec:kin}), with the one of BSSs, having larger masses, on average $\sim 1.2~\mathrm{M_{\odot}}$ (\citealt{raso19}).

\begin{figure}[!t]
\centering
\includegraphics[width=\textwidth]{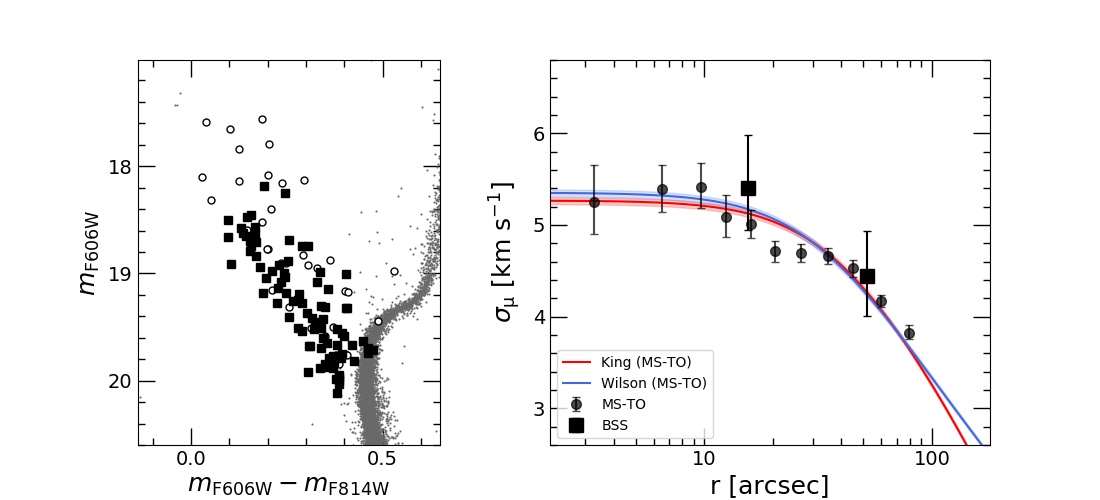}
\caption{Left panel: CMD of NGC~1261, zoomed in around the BSS region. We show as black squares the BSSs used for the kinematic analysis, and as empty circles those excluded because of too large PM errors. Right panel: BSS velocity dispersion profile (black squares). For comparison, we report the velocity dispersion profile of MS-TO stars and its best fit King and Wilson models (red and blue lines, respectively), already shown in Figure~\ref{fig:dispproffit}).}\label{fig:bsskin}
\end{figure}

Using the same technique presented in Section~\ref{sec:vdispprof}, here we obtain the combined velocity dispersion profile for BSSs in the plane of the sky.
As discussed above, most of the analyzed BSSs have PM errors larger than MS-TO stars. Indeed, the PM quality selection described in Section~\ref{ss:pmsel} causes a strong reduction of the BSS sample suitable for a kinematical study (see Figure~\ref{fig:mappm}). Quantitatively, we are left with only 32 BSSs, i.e.,  less than 25\% of the total sample. Moreover, due to the shape of the PM error distribution, this selection preferentially rejects bright BSSs, which correspond to the most massive ones (\citealt{raso19}). 
Therefore, applying this selection would imply a study limited only to BSSs with masses (and therefore kinematics) very similar to those of TO stars.
With the aim of mitigating this mass selection effect, and to maintain a large enough statistics, we thus dropped the (viii) selection criterion. With this selection, the BSS sample used for the kinematic analysis contains 76 stars, represented as black solid squares in Figure~\ref{fig:bsskin}. In this way, we cover almost the entire magnitude (hence, mass) extension of the BSS sequence, excluding only the brightest 0.5 mag, in which case the PM errors are too large to provide meaningful information about kinematics.

We divided the sample into two equally populated radial bins (38 stars each), and measured the velocity dispersion in each bin with the same method described in Section~\ref{sec:vdispprof}. The combined BSS velocity dispersion profile is shown in the right panel of Figure~\ref{fig:bsskin}. As apparent, it is consistent with the MS-TO velocity dispersion profile.
This is somehow expected from the derived value of $A^+_{\mathrm{rh}}$, which suggests a low level of dynamical evolution. Indeed, \citet{baldwin16} measured the  ratio between the BSS and the MS-TO velocity dispersions ($\alpha=\sigma_{\mathrm{BSS}}/\sigma_{\mathrm{MS-TO}}$) in 19 GCs and \citet{ferraro18treasury} compared $\alpha$ with the dynamical ages estimated through the $A^+$ parameter for 14 GCs in common between the two samples. They found an anticorrelation between the two parameters, where clusters with lower values of $A^+$ (i.e., less dynamically evolved) show larger values of $\alpha$ (see Figure 11 in \citealt{ferraro18treasury}). 
By inserting the value of $A^+_{\mathrm{rh}}$ determined above for NGC~1261 into Equation 3 of \citet{ferraro18treasury}, we expect $\alpha = 0.96 \pm 0.04$, which is fully compatible with what found from the comparison between the BSS and the TO velocity dispersion profiles. This result further confirms that NGC~1261 is poorly dynamically evolved.

\section{Summary and conclusions}\label{sec:conclu}

We constructed a high-precision \textit{HST} astro-photometric catalog of the central regions of NGC~1261, a relatively distant (D$_{\odot}$=15.7~kpc) Galactic GC with intermediate metallicity and low extinction ([Fe/H]=$-$1.27 and E($B-V$)=0.01, \citealt{h96}).

We determined the center of gravity and density profile of the system from resolved star counts, also complementing the \textit{HST} catalog with \textit{Gaia} DR2 data to cover the entire radial extension, out to $\sim 1$ deg. 
We found that the density profile shows a significant deviation from both a King and a Wilson profile in the outer regions (for $r>150^{\prime \prime}$ and $r>270^{\prime \prime}$, respectively). This confirms the existence of a diffuse stellar halo around the system, as previously shown by \citet{kuzma18}. 

We measured the \textit{HST} PMs of thousands of individual stars and used them to construct the velocity dispersion and anisotropy profiles in the plane of the sky of NGC~1261. We show that the central regions of the cluster (within r$\approx 80^{\prime \prime}$) are isotropic.  The central velocity dispersion from PMs is consistent with previous determinations from radial velocities.
Hence, under the assumption of global orbital isotropy, we combined the PM and LOS measurements and built the most extended velocity dispersion profile of NGC~1261 so far, covering a radial range from the very center, out to $r\sim800^{\prime \prime}$. Adopting the structural parameters estimated from the observed density profile and the central velocity dispersion obtained from PMs ($\sigma_0=5.3~\mathrm{km s^{-1}}$), we find a total mass of $\sim1.1 \times 10^5 ~\mathrm{M_{\odot}}$ for NGC~1261.

We also found that this GC hosts a quite large BSS population, composed of 132 stars. We measured the dynamical age of the system, by means of the parameter $A^+_{\mathrm{rh}}$ (\citealt{alessandrini16, lanzoni16, ferraro18treasury}), finding comparable results with previous estimates and confirming that it is relatively young from the point of view of dynamical evolution.
We also found a hint of a split BSS sequence observed by \citet{simunovic14}, although the detectability of the feature is critically dependent on binning.
The \textit{HST} PM measurements allowed us to also perform a kinematic study of the BSS population. In spite of large uncertainties due to small statistics, this confirms that NGC~1261 is in a moderate stage of dynamical evolution.

\acknowledgments

We thank the anonymous referee for useful comments that contributed to improve the presentation of the paper.
This paper is part of the project Cosmic-Lab (``Globular Clusters as Cosmic Laboratories'') at the Physics and Astronomy Department of the Bologna University (see the web page: \url{http://www.cosmic-lab.eu/Cosmic-Lab/Home.html}). The research is funded by the project \textit{Dark-on-Light} granted by MIUR through PRIN2017 contract (PI: Ferraro).
S.R. is grateful to A. Sollima for useful discussions.
A.B. and M.L. acknowledge support from STScI grant GO-13297 and GO-15232.
Based on observations with the NASA/ESA \textit{Hubble Space Telescope}, obtained at the Space Telescope Science Institute, which is operated by AURA, Inc., under NASA contract NAS 5-26555. 
This work has made use of data from the European Space Agency (ESA) mission Gaia (\url{https://www.cosmos.esa.int/gaia}), processed by the Gaia Data Processing and Analysis Consortium (DPAC, \url{https://www.cosmos.esa.int/web/gaia/dpac/consortium}).

\vspace{5mm}
\facilities{\textit{HST}(ACS/WFC; WFC3/UVIS)}
\software{\texttt{emcee} (\citealt{foremanmackey13}); 
\texttt{Matplotlib} (\citealt{matplotlibref}); 
\texttt{NumPy} (\citealt{numpyref})
}.

\end{document}